# Shape Resonances in superconducting gaps in a 2DEG at oxide-oxide interface

**Antonio Bianconi** [1]* **Davide Innocenti** [1], **Antonio Valletta** [2], **Andrea Perali** [3]

[1] RICMASS, Rome International Center for Materials Science Superstripes,
Via dei Sabelli 119A, 00185 Rome, Italy

[2] Department of Physics, Institute for Microelectronics and Microsystems, IMM CNR,
Via del Fosso del Cavaliere 100, 00133 Roma, Italy

[3] Physics Unit, School of Pharmacy, University of Camerino, 62032 Camerino, Italy

E-mail: antonio.bianconi@ricmass.eu;

**Abstract.** In multiband superconductivity, the case where the single electron hopping between different Fermi surface spots of different symmetry is forbidden by selection rules is recently attracting a large interest. The focus is addressed to superconductivity made of multiple condensates with different symmetry where the chemical potential crosses a 2.5 Lifshitz transition. This can now be investigated experimentally by fine-tuning of the chemical potential in the range of tens meV around a band edge using gate voltage control. We discuss here the case of a superconducting two-dimensional electron gas (2DEG), at the interface between two insulating oxides confined within a slab of 5 nanometers thickness, where the electronic structure is made of subbands generated by quantum size effects. We obtain shape resonances in the superconducting gaps, characterisc gaps to $T_c$ ratios and the BCS-BEC crossover in the upper subband for different pairing strength in the shallow Fermi surface, pointing toward the best configurations for enhanced superconductivity in 2DEG.

## 1. Introduction

There is growing agreement that advances in quantitative protocols for material design of high temperature superconductors should be based on the control of a) lattice parameters, b) defects distribution and c) tuning of the chemical potential, in heterostructures at atomic limit [1-13]. Since 1993 [1] it has been proposed that it is possible to increase the superconducting critical temperature by selecting an optimal geometrical architecture of composite materials made of atomic layers (or wires or stripes) intercalated by nanoscale spacers. The interest is focusing on tuning the chemical potential near a band edge where "shape resonances" in superconducting gaps appear in multi-condensates superconductivity [1-13].

The "shape resonance" is a type of Fano resonance in the configuration interaction between open and closed scattering channels, well known in atomic and nuclear physics [14-17] which occurs in the pairing processes in multi-condensates superconductors by the tuning the chemical potential near a band edge, as it was first recognized by Blatt [18,19]. It was first shown in 1996 [3-5] that the "shape resonance" gives the maximum of the critical temperature where the chemical potential is tuned above the band edge so that the Fermi energy in the small Fermi surface is of the order of one-two times the the energy scale of the pairing interaction. Therefore the shape resonance in the superconducting gaps gives a maximum Tc where the interacting Fermi liquid in the upper band is in the crossover regime from a Bose-like to a Fermi-like behavior [20-25,10].

---

* To whom any correspondence should be addressed.

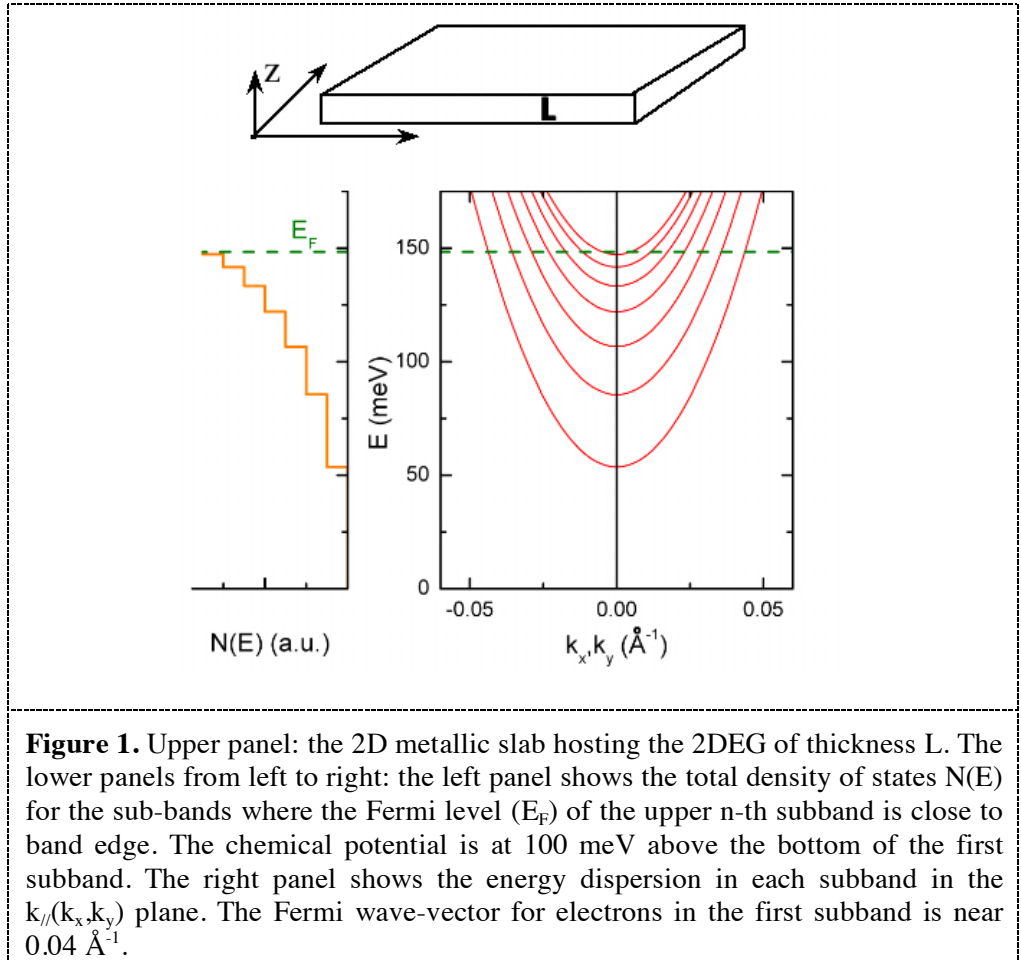

**Figure 1.** Upper panel: the 2D metallic slab hosting the 2DEG of thickness L. The lower panels from left to right: the left panel shows the total density of states N(E) for the sub-bands where the Fermi level ($E_F$) of the upper n-th subband is close to band edge. The chemical potential is at 100 meV above the bottom of the first subband. The right panel shows the energy dispersion in each subband in the $k_{//}(k_x,k_y)$ plane. The Fermi wave-vector for electrons in the first subband is near 0.04 Å$^{-1}$.

Superconductivity in thin metallic films with a thickness at atomic limit has been shown to persist to thicknesses much below the superconducting coherence length and even down to a single layer films [26-32]. Interestingly, in low dimensional superconductors, the superconducting coherence length (ξ) or the magnetic penetration depth (λ) do not define the critical length scale for the destruction of the superconducting state.

Layered complex metal oxides made of different modules provide systems with unique electronic, magnetic, ferroelectric and superconducting properties due to lattice heterogeneneity, valence fluctuations, lattice misfit strain, photoinduced effects, complex phase separation high temperature superconductivity [33-43].

At interfaces between complex oxides, electronic systems with unusual electronic properties can be generated [44,45]. Evidence of a two-dimensional electron gas (2DEG) at interfaces between two oxides $LaAlO_3$ and $SrTiO_3$ has been reported by several groups [46-49]. This 2DEG provides a superconducting layer that is a realization of a two-dimensional superconductor in a thin slab of finite thickness with superconducting transition temperature of 200 milli-Kelvin [45-48].

Maevasana et al. [49] have found a 2DEG with electron density as large as 8x10$^{13}$ cm$^{-2}$ formed at the bare $SrTiO_3$ surface of a cleaved crystal and its density can be controlled through exposure of the surface to intense ultraviolet light. The results show that the chemical potential can be tuned above the bottom of the second sub-band due to quantum size effects. These findings have been confirmed by a systematic study using angle-resolved photoemission spectroscopy (ARPES) providing new insights into the electronic structure of the 2DEG [50]. They shed light on previous observations in $SrTiO_3$-based heterostructures and suggest that different forms of electron confinement at the surface of $SrTiO_3$ lead to essentially the same 2DEG. There is therefore compelling evidence for an exotic 2DEG that forms at oxide interfaces based on $SrTiO_3$, but its precise nature remains elusive.

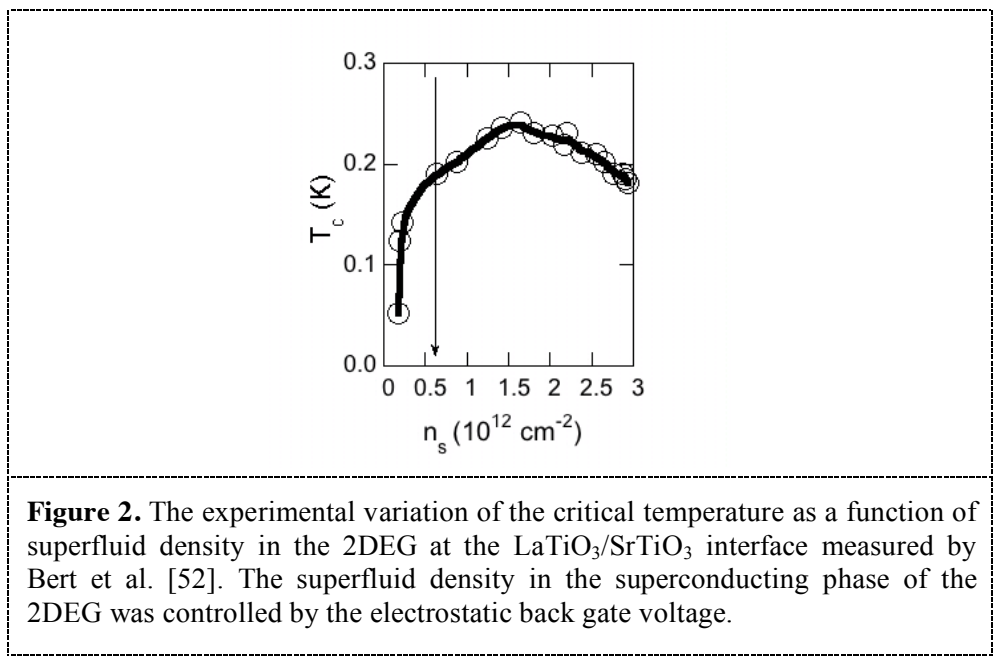

**Figure 2.** The experimental variation of the critical temperature as a function of superfluid density in the 2DEG at the $LaTiO_3/SrTiO_3$ interface measured by Bert et al. [52]. The superfluid density in the superconducting phase of the 2DEG was controlled by the electrostatic back gate voltage.

Recently Biscaras et al. [51] and Bert et al. [52] have investigated the superconducting phase by controlling the chemical potential in the 2DEG by electrostatic back gate voltage. Biscaras et al. have shown that the superconducting 2DEG is confined into a slab of about 5 nanometers at the $LaTiO_3$ side of the (Mott insulator)/$SrTiO_3$ (band insulator) interface forming subbands due to quantum size effect like it is shown in Figure 1.

Figure 1 shows the 2D metallic slab hosting the 2DEG of thickness L. The total density of states N(E) for the two dimensional subbands made of sharp steps at the band edge of each subband controlled by quantum size effects. By using the gate voltage the chemical potential is tuned to the bottom of the upper subband at about 100 meV above the bottom of the lowest subband. The subbands show a dispersion in the $k_{//}(k_x,k_y)$ plane. The electronic occupation of the upper subband controls the superconducting properties via the small Fermi surface produced by electrons in the upper n-th subband.

The chemical potential in the 2DEG is tuned by the gate voltage to the point where it crosses the bottom of the upper n-th subband, i.e., a 2.5 Lifshitz transition for appearing of a new Fermi surface spot near $k_{//}(k_x,k_y)=0$. Changing the gate voltage field the superconducting transition temperature shows a sharp increase as soon as the small Fermi surface spot due to finite Fermi energy in the n-th subband appears. The critical temperature increases to a maximum value followed by a smooth decrease as a function of the gate voltage [51]. Bert et al. [52] has reported magnetization and susceptibility measurements made by using a scanning SQUID (Superconducting Quantum Interference Device) [52], with a 3 μm diameter pick-up loop and a concentric field coil for applying a local AC magnetic field. They have been able to measure the superfluid density of the superconducting state and to obtain the curve of the critical temperature as a function of the superfluid density shown in Figure 2.

## 2. Results and discussion

We have calculated the superconducting gaps and the critical temperature in a metallic slab of 5 nanometers using our theoretical approach described in our previous papers [10,11] which is able to describe the superconducing phase near a band edge in a multiband system. The electrons are confined by a very high barrier of 5 eV within the slab and the Schrödinger equation is solved to get the wave-function of electrons in different Fermi surface spots. The electronic structure of the normal phase is made of multiple subbands as shown in Figure 1. The chemical potential is tuned around the bottom of the 6$^{th}$ subband in an energy range of meV. The BCS multi gaps equations for multiband superconductivity are calculated together with the density equation. The chemical potential $\mu$ is tuned around the bottom of the upper subband E of n-th subband (n=6). In the band edge region of the n-subband there is a strong renormalization of the chemical potential going from the normal to the superconducting phase which should be included in any theoretical approach.

The superconducting gaps in the subbands of the 2DEG in the slab of 5 nanometer thickness at the $LaTiO_3/SrTiO_3$ interface are shown in Figure 3. The chemical potential is tuned by changing the charge density form the bottom of the first subband. At each 2.5 Lifshitz transition for appearing of a new Fermi surface spot at the band edge of the n-th subband a new condensate is formed and its gap energy increases from zero up to reach a maximum. The superconducting gaps in the n-1 subbands with m<n are very similar. The figure shows three cases a) intraband coupling in the m-th subbands (m<n) in the very weak coupling range $c_{mm}=0.1$. b) intraband coupling in the n-th subband in the strong coupling range $c_{nn}=0.5$, with an interband $c_{n,i}=0.1$ c) the intraband coupling is the same as in panel b but now there is a strong exchange pair transfer term $c_{ni}=0.5$

The position of the chemical potential is measured by the Lifshitz parameter $z=(\mu-E_{n,L})/\omega_o$ where $E_{n,L}$ is the energy of the bottom of the n-th electronic subband and $\omega_o$ is the energy cut off of the pairing interaction. When the Lifshitz parameter z is in the range -1<z<0 the n-th subband is empty in the normal state but a Bose-like condensate is formed there by bosons created by the strong pairing interaction in the upper n-th subband and the pair transfer from the lower m-th subbands. For 0<z<1 the Fermi liquid in the n-th subband forms a small Fermi surface where the Fermi energy is smaller than the pairing interaction energy. Therefore the electrons in the n-th subband are in the anti-adiabatic regime where polaron formation is expected and in the superconducting phase bipolarons could form and condense. In this regime the wave-vector of electrons in the n-th Fermi surface $k_{Fn}$ could be smaller than the inverse of the superconducting coherence length $1/\xi_o$ where $(k_{Fn}\,\xi_o)<2\pi$ and $T_c$ versus superfluid density follows the Uemura plot. In cuprates this regime is called the "underdoped regime" or "pseudogap regime". In the range 1<z<2 the superconducting phase is in a crossover regime approaching the BCS regime for z>2.

We report in Figure 3 the calculation of the superconducting gaps in the upper subband n-th and in the (n-1)-th subband and the critical temperature. We assume a weak coupling regime in the lower subbands taking for the intraband coupling strength the value c=0.1.

The green region in Figure 3 corresponds to values of the Lifshitz parameter z, (-1<z<0) therefore the n-th subband is empty with no associated Fermi surface in the normal phase. On the other hand, the effective pairing interaction centered around the chemical potential is already able to make available for pair formation states in the n-th subband, thanks to its extension in energy of the order of $\omega_0$ induced by the retardation effects of the electron-phonon coupling. In this situation the condensate component of the n-th subband starts to form well before the appearing of a new Fermi surface in the normal state. This condensate has a Bose-like character, being the underlying Fermi surface not yet formed and its coherence factor $v(k)^2$ quite delocalized as a function k, implying Cooper pairs with a local character in real space.

The yellow region in Figure 3 corresponds to positive values of the Lifshitz parameter z, $0<z<1$: the chemical potential is above the bottom of the n-th subband and a new (small) Fermi surface appears in the electronic spectrum. The superconducting gap is of the order of the energy difference $\mu-E_{n,L}$ (which is the Fermi energy for the n-th subband) and the system is in the crossover regime between BCS and BEC.

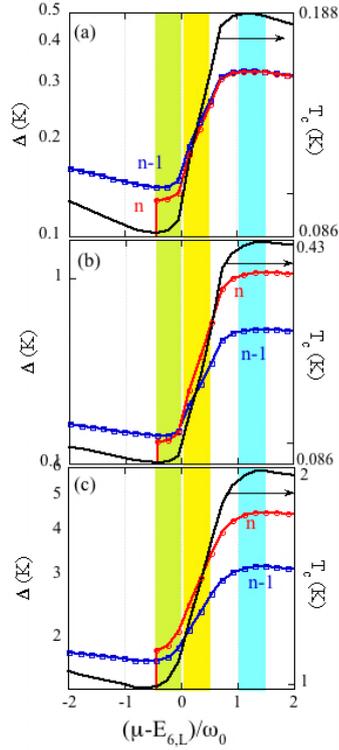
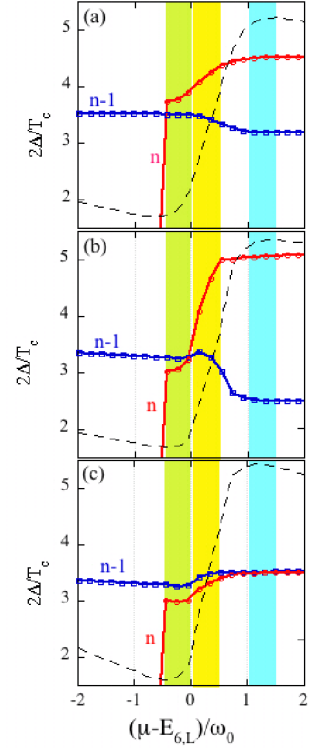

**Figure 3.** The superconducting gaps and the critical temperature as a function of the Lifshitz parameter $z=(\mu-E_{6,L})/\omega_o$. Panel (a): intraband and interband coupling in each subband $c_{nn}=c_{mm}=c_{nm}=0.1$; panel (b) the case of strong intraband coupling in the upper n=6 subband, $c_{nn}=0.5$, while $c_{mm}=c_{nm}=0.1$; c) with strong exchange pair transfer interband term $c_{nm}=c_{nn}=0.5$ while the intraband coupling in the lower subbands is $c_{nm}=0.1$.

**Figure 4.** The gap ratio $2\Delta_n/T_c$ in the upper subband and the gap ratio on the lower subbands $2\Delta_m/T_c$, $m<n$ as a function of the Lifshitz parameter $z=(\mu-E_{n,L})/\omega_o$, n=6, where m is the chemical potential and $E_{n,L}$ the energy of the bottom of the n-th subband. Panels (a), (b) and (c) show the results of calculations for the same corresponding cases as in Figure 3.

The precise extension of the crossover region depends from the couplings regime, being wider in the Lifshitz parameter z for stronger couplings, leading to larger gap values in the n-th subband, as in panel (c) of Figure 3. This crossover regime is highly non trivial, because the full superconducting system has a condensate which is a coherent mixture of BCS-like (forming in (n-1)-th subband) and Bose-like pairs (in the n-th subband) and the local Fermi energy $E_{Fn,L} = \mu-E_{n,L}$ is less or of the same order of the energy cutoff of the interaction $\omega_0$, which leads to non adiabatic effects in the superconducting properties in which the electron-phonon coupling play a role. For this reason this yellow region in Figure 3 can be also addressed as a crossover-antiadiabatic regime.

The blue region in Figure 3, with $1<z<1.5$, corresponds to a more conventional crossover to BCS regime for multi-band superconductivity. The frontier between the crossover regime and the BCS regime has the optimal characteristics for the amplification of superconductivity in 2DEG. Anti-adiabatic effects, which in general suppress superconductivity for conventional electron-phonon couplings, start to be negligible, while superconducting fluctuations can be lowered in their detrimental effects by the coexistence of small and large well formed Fermi surface in the spectrum, with the large Cooper pairs forming in the large Fermi surface being able to transfer their large superconducting stiffness to the smaller Copper pairs of the small Fermi surface by means of the interband Josephson-like exchange coupling. Finally, for $z>2$, the 2DEG superconducting system is in a BCS regime of conventional multi-band superconductivity.

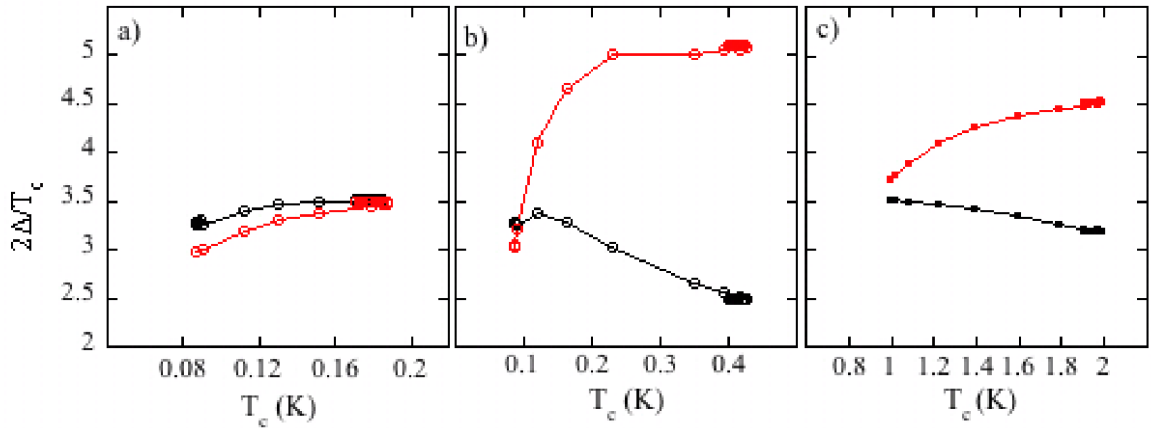

**Figure 5:** The gap ratio in the n-th (red dots) and m-th subbands (black dots) versus the critical temperature tuning the chemical potential in the 3 considered cases of figures 3 and 4. The figures show three cases a) intraband and interband coupling in all subbands are the weak coupling range 0.1; b) the intraband coupling in the upper n-th subband in the strong coupling range 0.5; c) intraband coupling in the n-th (n=6) and m-th subbands are like in panel (b) but the exchange pair transfer term is much larger, $c_{nm}=0.5$.

In Figure 5 we report the variation of the gap ratio $2\Delta/T_c$ as a function of the superconducting critical temperature. We show that where the intraband pairing terms are weak as well as the interband pair transfer interaction in panel (a) the gap ratio $2\Delta/T_c$ is similar in different bands, and it is close to the BCS standard value 3.5 for a single large Fermi surface. In panel 2 the BCS ratio in the upper subband, which is in the strong electron-phonon coupling limit, is much larger than 3.5 while for the other subbands, which are in the weak coupling limit $c_{mm}=0.1$, is much smaller. Panel (c) shows that by increasing the interband pair transfer term $c_{nm}=0.5$ the difference of the gap ratio in different bands decreases and it approaches again the BCS ratio 3.5.

In Figure 6 we plot the critical temperature as a function of the superfluid density for the three cases we have investigated. The results clearly show that in the mixed Bose-Fermi regime the critical temperature increases as a function superfluid density like in the Uemura plot and by reaching the Fermi regime and the multiband BCS condensation regime the critical temperature saturates and decreases increasing the superfluid density for all investgated cases as in the low superfluid density regime in Figure 2. Further work is needed to clarify the strength of the pair transfer term between different electronic compontens at the Fermi level in this 2DEG.

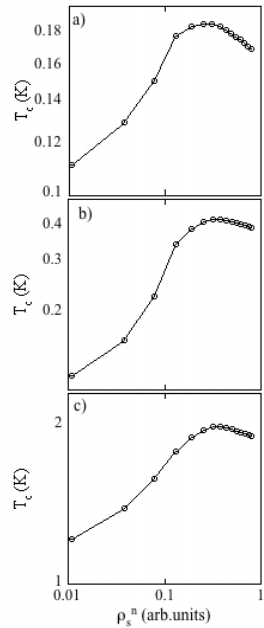

**Figure 6:** The critical temperature versus the superfluid density in the new appearing subband in the log-log plot. The figure shows three cases (panel a) intraband coupling in the n-th subband in the very weak coupling range 0.1; (panel b) intraband coupling in the n-th subband in the strong coupling range 0.5, (panel c) where the intraband coupling parameters are like in the panel (b) but with a much stronger exchange pair transfer term $c_{nm}=0.5$.

## 3. Conclusions

In conclusion one of our most important results: the superconducting shape resonance leading to the optimal amplification of the gap energies and of the critical temperature is located in the range $1<z<2$, and the critical temperature and the gaps in the lower m subbands have minima in the rang $-1<z<0$. It is indeed necessary for the system to go over the Lifshitz critical point to stabilize at the same time lattice and superconducting fluctuations. Optimal superconductivity therefore needs two formed and underlying Fermi surfaces: one large with weak coupling and another smaller one with strong coupling, with the local Fermi level close to the band bottom of the upper band, at an energy distance of the order of the characteristic energy scale of the pairing interaction, together with the interband exchange Josephson-like coupling allowing for (constructive) interference effects between the condensate of the bands.